# Total lineshape analysis of α-tetrahydrofuroic acid ¹H NMR spectra

Artemiy I. Nichugovskiy[1], Dmitry A. Cheshkov[2*]

[1]*Lomonosov Institute of Fine Chemical Technologies, MIREA - Russian Technological University, 86 Vernadsky Ave., 119571 Moscow, Russia*
[2]*State Scientific Research Institute of Chemistry and Technology of Organoelement Compounds, 38 Shosse Entuziastov, 105118 Moscow, Russia*

Moscow, 2022.

**Graphical abstract**

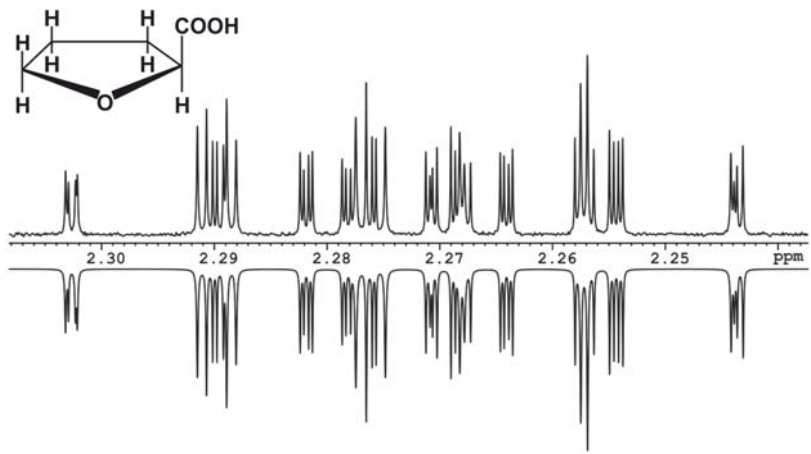

**Abstract**

The ¹H NMR spectra of L-proline oxygen analogous, α-tetrahydrofuroic acid, 7-spins nonsymmetric spin system with strongly pronounced non-first order effects were analyzed by ANATOLIA total lineshape fitting. A close agreement of theoretical and experimental spectra (with R-Factor values bellow 5%) was achieved and accurate values of coupling constant were obtained. The carboxylic substituent disrupts tetrahydrofuran symmetry and allows to unambiguously determine the typical values of geminal coupling constants in oxygen-containing saturated five-membered ring systems.

**Introduction**

Quite a large number of studies have been devoted to the investigation of the structural features of saturated five-membered heterocyclic compounds with one O, N, S heteroatom and its derivatives such as tetrahydrofuran[1,2,3], pyrrolidine[4,5], tetrahydrothiophene[6,7], L-proline[8,9]. The essential supplier of structural information is the data extracted from high-resolution NMR spectra, mainly the values of chemical shifts and spin-spin coupling constants.

However, the spectra of such compounds often reveal strong non-first-order effects, and the parameters of spin systems cannot be easily determined and required sophisticated techniques such as total lineshape fitting. Two pitfalls can be identified: one is associated with the complexity of spectra fine structure, especially for asymmetric spin systems with many non-zero coupling constants and close values of chemical shifts. The penalty function for such spectra possesses large dimensionality with plenty of local minima (as we have previously reported[10], incorrect data sometimes even fall into the literature, which definitely indicates the complexity of the spectra analysis process in such cases). Another pitfall has a fundamental nature of interconnection of spin system parameters with NMR spectrum and the ability of parameters determination from the spectrum. The simplest case is the absence of impact of coupling constants between magnetically equivalent spins on the spectrum. However, sometimes coupling constants cannot be unambiguously determined even for magnetically non-equivalent spins. For example, for $C_{2v}$-symmetric $[A']_4[X']_4$ spin systems of tetrahydrofuran, pyrrolidine and tetrahydrothiophene with not largely enough different $J_{AX}$ and $J_{AX'}$ coupling constants, the separate determination of geminal coupling constants ($^2J_{AA'}$, $^2J_{XX'}$) from NMR spectrum are impossible[1,6,11]. This situation is very similar to the case of AA'XX' spin system with values of $J_{AX}$ and $J_{AX'}$ approaching each other, which leads the entire spin system approaches $A_2X_2$ type with vanishing of influence of $J_{AA'}$ and $J_{XX'}$ coupling constants on NMR spectrum. One possible way to overcome this situation is a compound derivatization with introduction of asymmetry to the spin system and therefore chemical non-equivalence of previously equivalent spins. In the case of tetrahydrofuran, we have decided to perform thorough study of α-tetrahydrofuroic acid $^1$H NMR spectra, not only due to its asymmetric spin system type, but also due to structural similarity to the natural amino acid L-proline.

**Analysis of 300 MHz $^1$H NMR spectrum (at 303K) of 0.5M α-tetrahydrofuroic acid methanol-$d_4$ solution**

The spectrum analysis begins with generation of trial parameters set. First, for this purpose, all signals should be assigned to the certain protons in the molecule. Usually, the signal assignment is based on chemical shifts, coupling constant values and cross-peaks presented in 2D correlational spectra such as COSY, HSQC, HMBC and NOESY. However, the case of α-tetrahydrofuroic acid is very similar to L-proline, where the resonant multiplets possess complicated fine structure due to non-first order effects. Also, the multiplets of methylene protons have almost the same total width, due to the presence of equal number of *cis*- and *trans*- vicinal coupling partners. Finally, the small molecular size and strong non-first order coupling effects makes the usage of 2D technics like NOESY useless. Earlier we have performed thorough analysis of L-proline $^1$H spectra with the involvement of Monte-Carlo optimization method (simulated annealing) for the overcoming of such ambiguities[10]. Figure 1 represents the superposition of L-proline and α-tetrahydrofuroic acid spectra. Given the similarity of these spectra, we decided to check whether the order of the signals in the α-tetrahydrofuroic acid and L-proline coincides. The resonant frequencies were approximately set to multiplet gravity centers, the values of vicinal coupling constants were set as for tetrahydrofuran (taken from [1,11]), the value of geminal coupling constant for α-CH$_2$ protons was set to -8 Hz and for β-CH$_2$ protons was set to -12 Hz. The $^4J$ long-range coupling constants were set to 0 Hz except for $^4J_{trans-\alpha\beta}$, which were set to -0.5 Hz. Spectral linewidth (0.2 Hz) was approximately determined from a linewidth measurement of single multiplet lines. The full set of initial parameters is present in table 1. For the total lineshape analysis of the $^1$H NMR spectra we used developed in our group program ANATOLIA (v1.2)[12] which is based on the effect of local minima elimination by application of additional Lorentzian broadening to both experimental and theoretical spectra[11-13]. Additional broadening is gradually decreased in the course of the spectrum analysis.

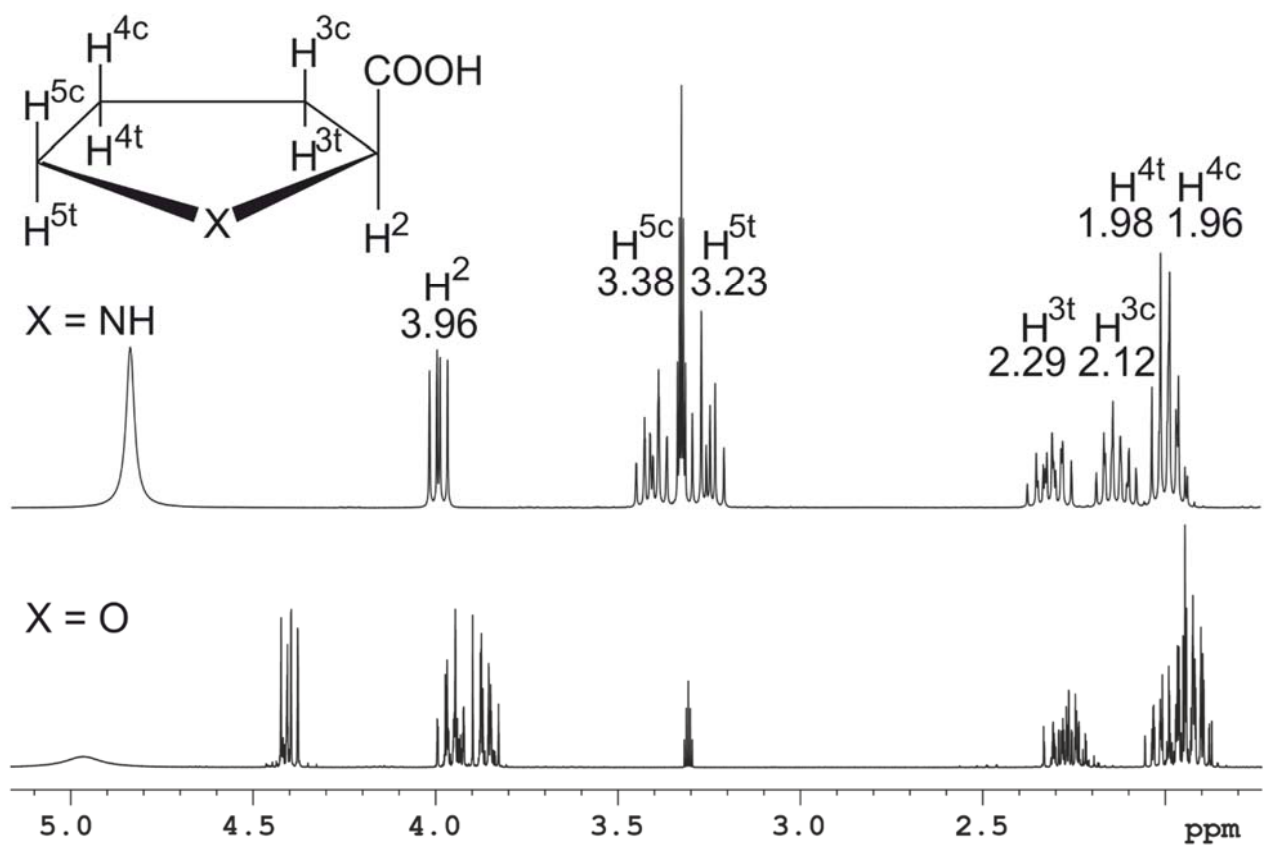

Figure 1. $^1$H NMR spectra (300 MHz, 303K) of L-proline and α-tetrahydrofuroic acid methanol-$d_4$ solutions.

First optimization run was performed with relatively large additional line broadening sequence of 12 10 8 6 4 3 Hz, with the optimization of resonant frequencies and spectrum magnitude only. The second run was performed with single broadening of 3 Hz starting from results of the first run with additionally defrosted geminal and vicinal coupling constants and spectral linewidth. The results of two sequential optimization runs are represented by table 2 and figure 2. Briefly, already on this preliminary optimization stage the adequate correspondence of broadened (LB = 3 Hz) experimental and theoretical spectrum is achieved (R-Factor is 1.75%). It should be noted that the values of resonant frequencies of the two most upfield shifted protons ($H^{4c}$ and $H^{4t}$, which also have closest values of resonant frequencies) were switched during the first optimization run. Therefore, the sequence of signals in the α-tetrahydrofuroic acid slightly differs from that of L-proline. However, the difference in the chemical shifts of these protons is only 0.016ppm (5.03 Hz on 300 MHz spectrometer). Taken as a whole, the result of two sequential optimizations can be considered as a moderate change in the initial parameters and indicates the correctness of the approach to the formation of the initial parameter set based on the tetrahydrofuran spin-spin coupling constants and L-proline chemical shifts, which allowed to avoid falling into local minima.

The third optimization run was performed with all 30 parameters unfixed (7 for resonant frequencies, 21 for coupling constants, two for linewidth and spectrum magnitude) with broadening sequence of 3 2 1 0.5 0.2 0.1 0 Hz. The resultant parameters set presented in table 2, figure 3 and characterized by R-Factor of 4.8%.

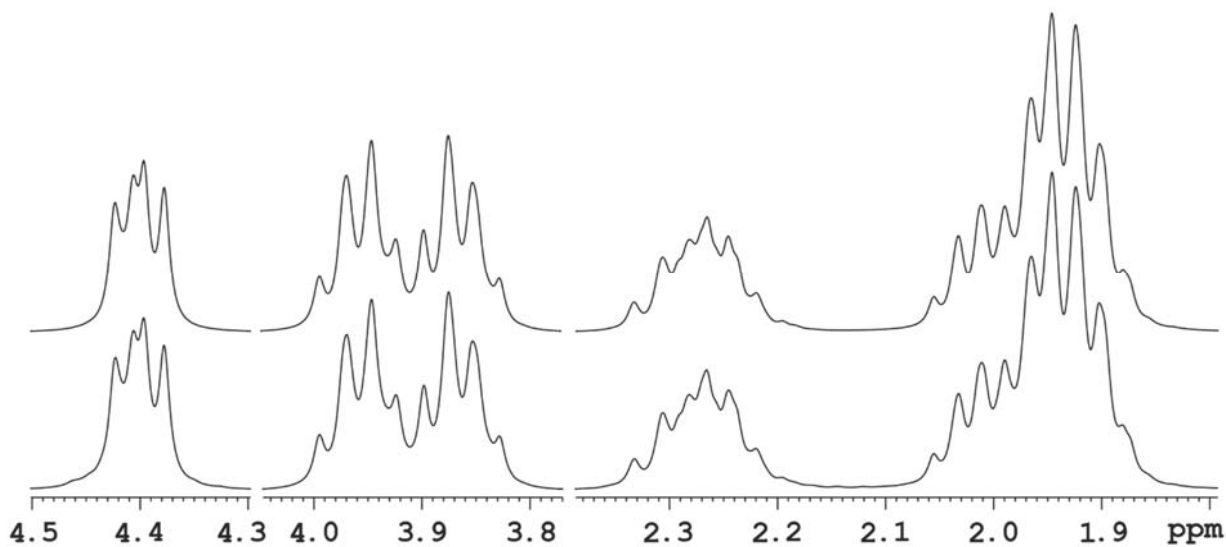

Figure 2. The result of preliminary optimization (after two sequential runs) of resonant frequencies, geminal and vicinal coupling constants, linewidth and spectrum magnitude (top) for experimental 300 MHz $^1$H NMR spectrum (303K) of 0.5M α-tetrahydrofuroic acid methanol-$d_4$ solution broadened with LB of 3 Hz (bottom).

Table 1. First trial parameters set and preliminary optimization results for broadened (LB = 3 Hz) experimental 300 MHz $^1$H NMR spectrum (303K) of 0.5M α-tetrahydrofuroic acid methanol-$d_4$ solution.

| No. | Parameter | Value(ppm/Hz) | Prelim. opt. | No. | Parameter | Value(Hz) | Prelim.opt. |
|---|---|---|---|---|---|---|---|
| 1 | $\nu_{1(2)}$ | 4.40/1321.0 | 4.40/1320.80 | 16 | $J_{25}$ | 0.0 | 0.00 |
| 2 | $\nu_{2(5c?)}$ | 3.95/1185.0 | 3.96/1187.58 | 17 | $J_{26}$ | 6.1 | 6.26 |
| 3 | $\nu_{3(5t?)}$ | 3.87/1160.0 | 3.87/1159.88 | 18 | $J_{27}$ | 7.6 | 7.44 |
| 4 | $\nu_{4(3t?)}$ | 2.27/680.0 | 2.27/681.78 | 19 | $J_{34}$ | 0.0 | 0.00 |
| 5 | $\nu_{5(3c?)}$ | 1.99/597.0 | 2.00/600.15 | 20 | $J_{35}$ | -0.5 | -0.50 |
| 6 | $\nu_{6(4t?)}$ | 1.95/585.0 | 1.92/575.51 | 21 | $J_{36}$ | 7.6 | 7.44 |
| 7 | $\nu_{7(4c?)}$ | 1.93/578.0 | 1.93/580.02 | 22 | $J_{37}$ | 6.1 | 6.05 |
| 8 | $J_{12}$ | 0.0 | 0.00 | 23 | $J_{45}$ | -12.0 | -12.68 |
| 9 | $J_{13}$ | 0.0 | 0.00 | 24 | $J_{46}$ | 8.7 | 8.74 |
| 10 | $J_{14}$ | 7.6 | 8.48 | 25 | $J_{47}$ | 6.1 | 6.21 |
| 11 | $J_{15}$ | 6.1 | 5.77 | 26 | $J_{56}$ | 6.1 | 6.14 |
| 12 | $J_{16}$ | 0.0 | 0.00 | 27 | $J_{57}$ | 8.7 | 8.35 |
| 13 | $J_{17}$ | -0.5 | -0.50 | 28 | $J_{67}$ | -12.0 | -12.38 |
| 14 | $J_{23}$ | -8.0 | -8.10 | 29 | LW | 0.2 | 0.35(+3Hz) |
| 15 | $J_{24}$ | -0.5 | -0.50 |  | R-Factor | - | 1.75% |

**Clarification of $^4J$ long-range coupling constants relative signs.**

For the accurate determination of long-range coupling constant relative signs, we additionally performed 256 screening calculations with total signs alternation for eight $^4J$ coupling constants ($2^8 = 256$) starting from the obtained $^4J$ absolute values and without application of any additional broadening. Surprisingly, all these 256 starting parameter sets after optimization fall into the same minimum obtained earlier.

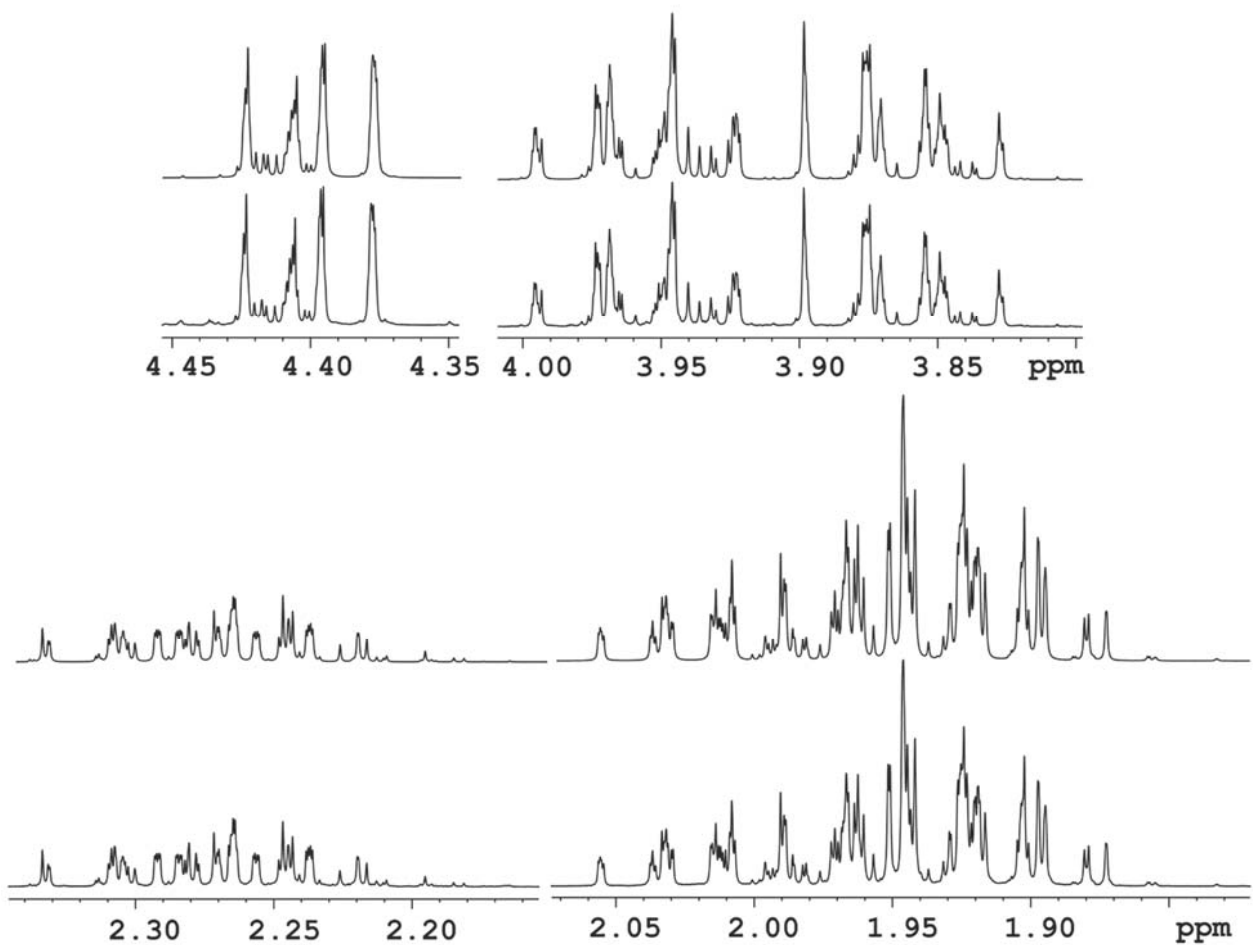

Figure 3. The result of the total lineshape analysis (top) for experimental 300 MHz $^1$H NMR spectrum (303K) of 0.5M α-tetrahydrofuroic acid methanol-$d_4$ solution (bottom).

**Analysis of 600 MHz $^1$H NMR spectrum (at 303K) of 0.5M α-tetrahydrofuroic acid methanol-$d_4$ solution**

Also, we performed measurement and analysis of the spectrum for the same sample recorded at 303K on 600 MHz spectrometer. The resonant frequencies for the initial parameter set were approximately determined as multiplet gravity centers, and the coupling constants values were taken from the result of 300 MHz spectrum analysis. The spectrum analysis in this case consisted of two successive optimization runs. During the first run only resonant frequencies, spectrum linewidth and magnitude were optimized with broadening sequence of 12 10 8 6 4 3 Hz. During the second run, all 30 parameters were allowed to vary and broadening sequence of 3 2 1 0.5 0.2 0.1 0 Hz were applied. The obtained optimization result (table 2, figure 4) is characterized by the R-Factor of 4.24% and closed coupling constants values to the 300 MHz solution.

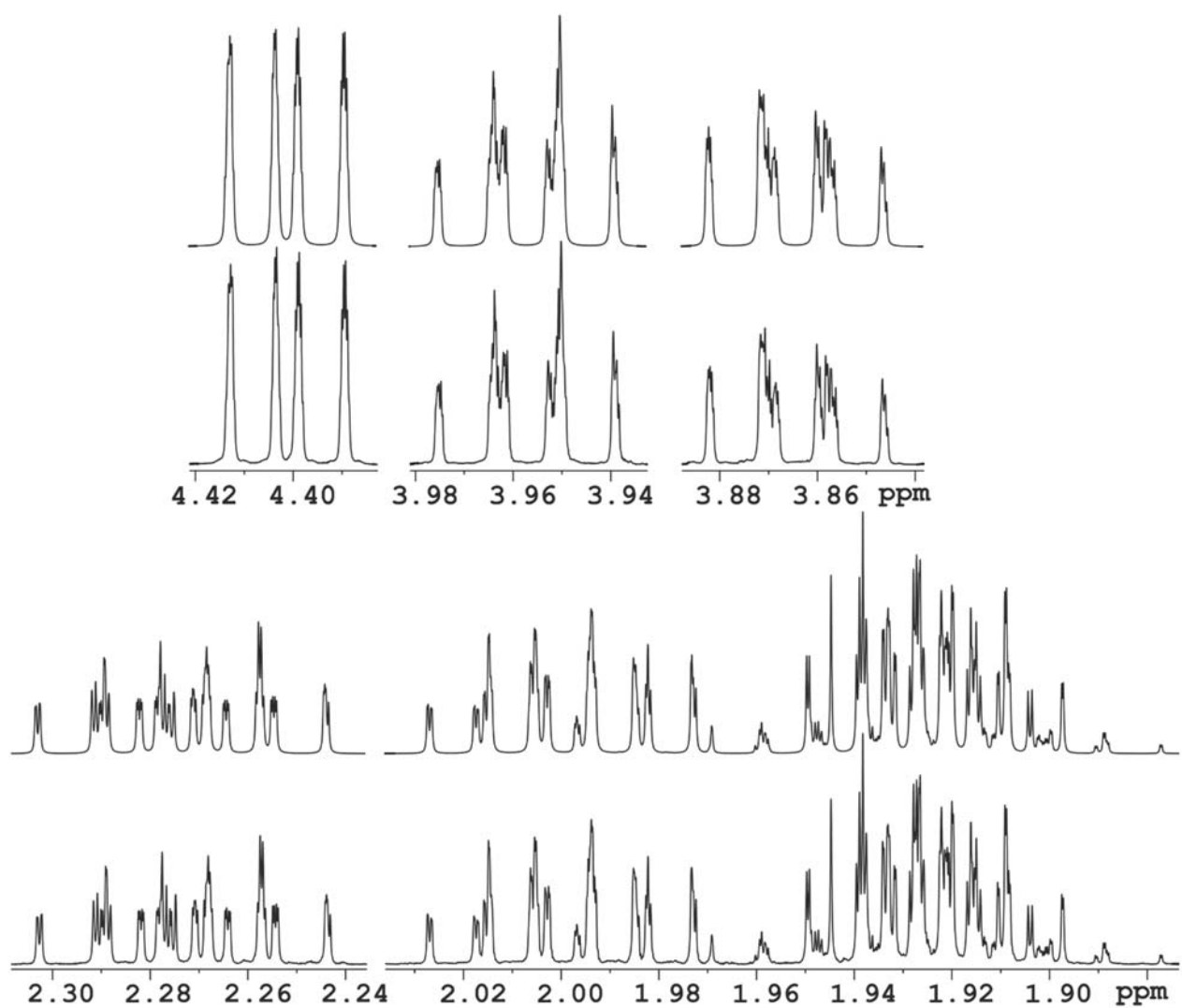

Figure 4. The result of the total lineshape analysis (top) for experimental 600 MHz $^1$H NMR spectrum (303K) of 0.5M α-tetrahydrofuroic acid methanol-$d_4$ solution (bottom).

**Conclusion**

In the present work the relatively complicated $^1$H NMR spectra of α-tetrahydrofuroic acid methanol-$d_4$ solution, 7-spins nonsymmetric spin system with strongly pronounced non-first order effects were thoroughly analyzed by a total lineshape fitting. The obtained accurate and precise values of spin-spin coupling constants are not only of fundamental importance (since they expand our knowledge about spin-spin coupling constants in saturated five-membered rings) but also can be used further for the investigation of conformational equilibrium and pseudo-rotation process in α-tetrahydrofuroic acid.

**Experimental section. Sample preparation, magnet shimming and spectra processing**

For the recording of high-quality NMR spectra, a precise magnet shimming with the involvement of orthogonal shimming algorithm proposed by C.A. Michal[14], which allows to obtain the maximal possible field homogeneity on the certain spectrometer was used. Spectra were recorded on Bruker AVANCE 600 and Bruker DPX 300 spectrometers after a thorough

Table 2. Results of the total lineshape analysis for $^1$H NMR spectra of α-tetrahydrofuroic acid.

| No. | Parameter | 300 MHz | 600 MHz | No. | Parameter | 300 MHz | 600 MHz |
|---|---|---|---|---|---|---|---|
| 1 | $\nu_{1(2)}$ | 4.40/1320.779 | 4.40/2640.821 | 16 | $J_{25}$ ($^4J_{3c,5c}$) | 0.091 | 0.093 |
| 2 | $\nu_{2(5c)}$ | 3.96/1187.636 | 3.96/2374.065 | 17 | $J_{26}$ ($^3J_{4t,5c}$) | 6.400 | 6.400 |
| 3 | $\nu_{3(5t)}$ | 3.87/1159.880 | 3.87/2318.882 | 18 | $J_{27}$ ($^3J_{4c,5c}$) | 7.258 | 7.254 |
| 4 | $\nu_{4(3t)}$ | 2.27/681.838 | 2.27/1363.503 | 19 | $J_{34}$ ($^4J_{3t,5t}$) | 0.192 | 0.190 |
| 5 | $\nu_{5(3c)}$ | 2.00/600.211 | 2.00/1199.527 | 20 | $J_{35}$ ($^4J_{3c,5t}$) | -0.518 | -0.517 |
| 6 | $\nu_{6(4t)}$ | 1.92/575.240 | 1.92/1149.995 | 21 | $J_{36}$ ($^3J_{4t,5t}$) | 7.294 | 7.296 |
| 7 | $\nu_{7(4c)}$ | 1.93/580.268 | 1.93/1159.759 | 22 | $J_{37}$ ($^3J_{4c,5t}$) | 6.102 | 6.100 |
| 8 | $J_{12}$ ($^4J_{2,5c}$) | -0.245 | -0.241 | 23 | $J_{45}$ ($^2J_{3c,3t}$) | -12.650 | -12.653 |
| 9 | $J_{13}$ ($^4J_{2,5t}$) | -0.235 | -0.235 | 24 | $J_{46}$ ($^3J_{3t,4t}$) | 8.287 | 8.284 |
| 10 | $J_{14}$ ($^3J_{2,3t}$) | 8.417 | 8.422 | 25 | $J_{47}$ ($^3J_{3t,4c}$) | 6.587 | 6.589 |
| 11 | $J_{15}$ ($^3J_{2,3c}$) | 5.747 | 5.752 | 26 | $J_{56}$ ($^3J_{3c,4t}$) | 6.388 | 6.397 |
| 12 | $J_{16}$ ($^4J_{2,4t}$) | -0.170 | -0.170 | 27 | $J_{57}$ ($^3J_{3c,4c}$) | 8.168 | 8.161 |
| 13 | $J_{17}$ ($^4J_{2,4c}$) | -0.556 | -0.560 | 28 | $J_{67}$ ($^2J_{4c,4t}$) | -12.167 | -12.170 |
| 14 | $J_{23}$ ($^2J_{5c,5t}$) | -8.114 | -8.115 | 29 | LW | 0.202 | 0.186 |
| 15 | $J_{24}$ ($^4J_{3t,5c}$) | -0.461 | -0.458 |  | R-Factor | 4.77% | 4.24% |

Standard deviations do not exceed 0.001 Hz for all parameters in both spectra.

preliminary shimming with optimization of all available shims. Prior to sample preparation the screening of NMR tubes was performed: for five high precision NMR tubes (Wilmad 535-PP) the $^1$H NMR spectra of 0.05M TMS solution in acetone-$d_6$ were recorded with sample spinning at 20 Hz, and the tube with narrowest TMS signal and highest level of $^2$H-LOCK signal was selected.

For preparation of the sample commercial α-tetrahydrofuroic acid (98+%, ACROS Organics, USA) was used without any additional purification. The solution of 0.5M α-tetrahydrofuroic acid and 0.05M TMS in methanol-$d_4$ was placed to the selected tube and subjected to 20 minutes of degassing in an ultrasonic bath with air displacement by argon.

Both spectra were recorded using the standard relaxation-pulse-acquisition pulse program with application of 30° excitation pulses. Acquisition parameters were (600/300 MHz): SW 7.0/6.5 ppm, AQ 15.6/19.2 s, relaxation delay 20/8 s, number of scans 8/8. The temperature of 303 K was maintained during spectral acquisition. TMS signal was used for chemical shifts referencing. The multiplet integral intensities in obtained spectra are in perfect agreement with chemical structure, which indicates the absence of signals saturation.

For spectra resolution enhancement with retention of Lorentzian lineshape, reference deconvolution[15,16] was applied during spectra processing. The TMS signal was used as reference, theoretical reference signal was calculated with linewidth of 0.12 Hz taking into account the $^{29}$Si satellites. The acquired FIDs were zero filled once and subjected to Fourier transformation. After that, phase and baseline corrections were applied to the spectrum, from which the reference region was withdrawn. The imaginary parts of the spectra required for deconvolution were restored using Hilbert transformation. The resulting linewidth of the signals was about 0.18 and 0.20 Hz for 600 and 300 MHz spectrum respectively.

**Acknowledgments**

The authors thank Dr. P.A. Solovyev for helpful discussion.